\documentclass[twocolumn,aps,prc,superscriptaddress,floatfix,showpacs]{revtex4}
\usepackage{epsfig,graphics}
\usepackage{graphicx}% Include figure files
\usepackage{dcolumn}% Align table columns on decimal point
\usepackage{bm}% bold math
\usepackage{amsmath}
\usepackage[usenames]{color}
\usepackage{ulem} %% fo

\draft
\usepackage{color}

\begin{document}
%\begin{CJK*}{GBK}{song}%%windows
%\begin{CJK*}{GB}{gbsn}%Mac or Linux

%\renewcommand {\baselinestretch 2}

\title{Hard photon flow  and photon-photon  correlation in intermediate energy heavy-ion collisions}

%Maybe there is a bug in revtex4-1 about "\author" or "\affiliation"
\author{ Y. G. Ma\footnote{ ygma@sinap.ac.cn}}
\affiliation{Shanghai Institute of Applied Physics, Chinese
Academy of Sciences, Shanghai 201800, China}

\author{G. H. Liu}
\affiliation{Shanghai Institute of Applied Physics,
Chinese Academy of Sciences, Shanghai 201800, China}

\author{ X. Z. Cai}
\affiliation{Shanghai Institute of Applied Physics, Chinese
Academy of Sciences, Shanghai 201800, China}
\author{ D. Q. Fang}
\affiliation{Shanghai Institute of Applied Physics, Chinese
Academy of Sciences, Shanghai 201800, China}
\author{W. Guo}
\affiliation{Shanghai Institute of Applied Physics, Chinese
Academy of Sciences, Shanghai 201800, China}
\author{W. Q. Shen}
\affiliation{Shanghai Institute of Applied Physics, Chinese
Academy of Sciences, Shanghai 201800, China}
\author{W. D. Tian}
\affiliation{Shanghai Institute of Applied Physics, Chinese
Academy of Sciences, Shanghai 201800, China}
\author{H. W. Wang}
\affiliation{Shanghai Institute of Applied Physics, Chinese
Academy of Sciences, Shanghai 201800, China}

\date{ \today}

\begin{abstract}

Hard photons emitted from  energetic heavy ion collisions are 
very interesting since they  do not experience nuclear interaction, and 
therefore they are  useful to explore properties of nuclear matter. We
investigated hard photon production and its properties in
intermediate energy heavy-ion collisions with  the help of the
Blotzmann-Uehling-Ulenbeck model. Two components of hard photons
are discussed: direct  and thermal. The positive
directed flow parameter and negative elliptic flow parameter of
direct photons are demonstrated and they are anti-correlated to
the flows of free protons. The dependencies of hard photon production
and anisotropic parameters on  impact parameter, beam energy,
nuclear equation of state and symmetry energy are also discussed.
Furthermore, we investigated the two-photon momentum correlation
function from which the space-time structure information of the photon source could be
extracted as well as the two-photon azimuthal correlation which could
provide another good method to determine the elliptic flow parameter
$v_{2}$ of direct hard photons.
\end{abstract}


\maketitle

\section{Introduction}

One of the main goals of intermediate energy heavy ion collision
(HIC)
 is to study properties of nuclear matter,
especially to determine   the nuclear Equation-of-State (EOS).
HIC  provides a unique means to compress nuclear matter to hot and
dense  phase within a laboratory environment. The pressures that
result from the high densities achieved during such collisions
strongly influence on the motion of ejected matter and provide the
sensitivity to the EOS. In
comparison with the conventional hadronic probes, photons
interacting only weakly through the electromagnetic force with the
nuclear medium are not subjected to distortions by the final state
(neither Coulomb nor strong) interactions and therefore photon
delivers an undistorted picture of the emitting source
\cite{schutz,cassing,Bona,nifenecker}.

Since the last two decades, many model calculations and
experimental facts
\cite{schutz,cassing,nifenecker,Bona,wada,schutz1,schutz2,martinez,enterria}
have indicated that in intermediate energy heavy-ion collisions
hard photons defined as photons with energies above the giant
dipole resonance domain, above 30 MeV in this paper, mainly
originate from incoherent proton-neutron bremsstrahlung
collisions, namely $p+n \rightarrow p+n+\gamma$. A nice review paper on hard photon was available \cite{Bona}.
These hard
photons are emitted from two distinct sources. The first and
dominant component denoted as direct hard photon (called as "direct photon" for short)
is associated with the
first-chance proton-neutron collisions in the initial phase of the
heavy-ion reaction. The second one originates from the secondary
proton-neutron collisions in the later stage of the reactions when
the di-nuclear system tends to be thermalized, accordingly called
as thermal hard photon (called as "thermal photon" for short). Because of their distinct emission sources,
direct photons and thermal photons can deliver thermodynamic
and dynamical  information of  hot and dense nuclear matter
formed during the various stages of the heavy-ion collisions.

In relativistic heavy ion collisions, photons are  also of very
interesting since they can be served as one of the potential signals
of quark-gluon plasma (QGP)  formation, eg., see
Ref.~\cite{Phenix,Liu,Long,WA}.  A  hot QGP radiates a large
amount of thermal photons, which dominate the spectra at small
transverse momenta, whereas hard processes in nucleon-nucleon
scatterings produce large momenta photons. Therefore photon
enhancement at low transverse momenta  could be seen as a QGP
signal, which has been observed at the BNL-Relativistic Heavy-Ion
Collider  \cite{Phenix}. Similar to the feature of photon
production in intermediate energy HIC, photons emitted from the
interior of the hot matter no longer  interact with the hadronic
medium, in contrast to hadronic observables.

 The paper is organized as follows. In Sec. II, the
simulation tool which we used is briefly introduced and the
calculation method of photon production is outlined. Sec. III
describes the classification of hard photons and definition of
anisotropic flow, and presents the results of the azimuthal asymmetry for
direct photons and free protons. In Sec. IV, we discuss the
different variables (impact parameter, beam energy and EOS)
dependences of the hard photon production and/or anisotropic flow
parameters. Sec. V gives the results and discussions on
two-photon correlation functions, namely momentum correlation
function, which is also called Hanbury-Brown and Twiss (HBT)
correlation function,  as well as azimuthal correlation function. Finally
the summary is given in Sec. VI.

\section{Brief introduction to the model and photon production}

\subsection{The Blotzmann-Uehling-Ulenbeck Equation}

The transport model is very useful for treating heavy-ion collision
dynamics and obtaining important information of nuclear matter.  In
intermediate energy heavy-ion collisions, the
Blotzmann-Uehling-Ulenbeck (BUU) model is an  extensively useful tool
\cite{Bauer}. The BUU equation takes both Pauli blocking and the mean field into
consideration, reads

\begin{align}
      & \frac{\partial f}{\partial t}+ v \cdot \nabla_r f - \nabla_r U
\cdot \nabla_p f  = \frac{4}{(2\pi)^3} \int d^3p_2 d^3p_3 d\Omega
\nonumber
\\ & \frac{d\sigma_{NN}}{d\Omega}V_{12}
 \times [f_3 f_4(1-f)(1-f_2) - f f_2(1-f_3)(1-f_4)] \nonumber
\\ & \delta^3(p+p_2-p_3-p_4).  \label{BUU}
                   \end{align}

It is solved with the method of Bertsch and Das Gupta
\cite{Bertsch}. In Eq.(~\ref{BUU}), $\frac{d\sigma_{NN}}{d\Omega}$
and $V_{12}$ are in-medium nucleon-nucleon cross section and
relative velocity for the colliding nucleons, respectively, and
$U$ is the mean field potential including the isospin-dependent
term:

\begin{equation}
  U(\rho,\tau_{z}) = a(\frac{\rho}{\rho_{0}}) +
  b(\frac{\rho}{\rho_{0}})^{\sigma} + C_{sym} \frac{(\rho_{n} -
    \rho_{p})}{\rho_{0}}\tau_{z},
\end{equation}
where $\rho_0$ is the normal nuclear matter density; $\rho$,
$\rho_n$, and $\rho_p$ are the nucleon, neutron and proton
densities, respectively; $\tau_z$ equals 1 or -1 for neutrons and
protons, respectively; The coefficients $a$, $b$ and $\sigma$ are
parameters for nuclear equation of state.  Three sets of mean
field parameters are used, namely the soft EOS with the
compressibility $K$ of 200 MeV ($a$= -356 MeV, $b$ = 303 MeV,
$\sigma$ = 7/6), and the semi-soft EOS with $K$ of 235 MeV ($a$ =
-218 MeV, $b$ = 164 MeV, $\sigma$ = 4/3), and the hard EOS with
$K$ of 380 MeV ($a$ = -124 MeV, $b$ = 70.5 MeV, $\sigma$ = 2).
$C_{sym}$ is the symmetry energy strength due to the density
difference of neutrons and protons in nuclear medium, which is
important for asymmetry nuclear matter (here $C_{sym} = 32$ MeV is
used), but it is trivial for the symmetric system studied in the
present work.

\subsection{Production cross sections of bremsstrahlung hard photon}

The BUU model was shown to be very successful in describing the bulk properties of
the reaction and nucleon emission in intermediate-energy heavy-ion collisions. In addition,
the proton-neutron bremsstrahlung photon can be simulated as well in the model. For
determining the  elementary
double-differential hard photon production cross sections on the
basis of individual proton-neutron bremsstrahlung, the hard-sphere
collision was adopted from Ref.~\cite{Jackson}, and modified as in
Ref.~\cite{Cassing2} to allow for energy conservation. The double
differential probability is given by
\begin{equation}
\frac{d^2\sigma^{elem}}{dE_{\gamma}d\Omega_{\gamma}} = \alpha_c
\frac{R^2}{12\pi}\frac{1}{E_{\gamma}}(2\beta_f^2+3sin^2\theta_{\gamma}\beta_i^2).
\label{ddcs}
\end{equation}
Here $R$ is the radius of the sphere, $\alpha_c$ is the fine
structure constant, $\beta_i$ and $\beta_f$ are the initial and
final velocity of the proton in the proton-neutron center of mass
system, and $\theta_{\gamma}$ is the  angle between incident
proton direction and photon emitting direction. More details for
the model can be found in Ref.~\cite{Bauer,GHLiu}.

\section{Azimuthal asymmetry of hard photons}

\subsection{Definitions of direct photons and thermal photons}

In the present work, we simulate the reaction of $^{40}$Ca + $^{40}$Ca
collisions in most cases. Sometimes $^{48}$Ca + $^{48}$Ca and Kr +
Ni are also simulated.

\begin{figure}
 \includegraphics[scale=0.43]{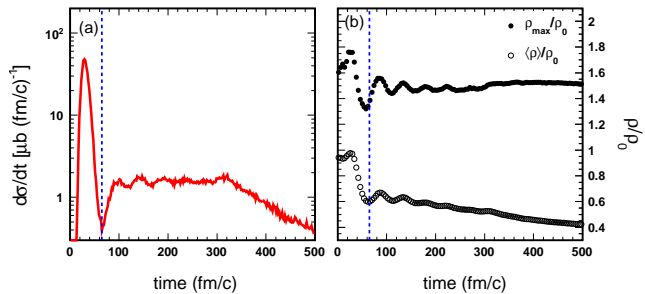}
 \vspace{0.1truein}
  \caption{\footnotesize  (Color online)  (a) Time evolution of hard photon production rate
  for the reaction  $^{40}$Ca + $^{40}$Ca at 60$A$ MeV in 40\%--60\%
  centrality. The EOS parameters with the compressibility $K$ of 235
  MeV are used;
    (b) Time evolution of reduced  maximum density $\rho_{max}/\rho_0$ (closed circles) and reduced average
    density $\langle \rho \rangle / \rho_0$ (open circles) of the whole reaction system in the same reaction.
    The blue dashed line represents the time when  till the first expansion stage, and in the panel (a) it separates
    direct hard photons (on the left side) and thermal hard photons (on the right side).
  }\label{time}
\end{figure}

Fig.~\ref{time}(a) and (b)  show the time evolutions of the production rate
of bremsstrahlung hard photons and of system densities, including
both maximum density (closed circles) and average density (open circles), respectively,
for $^{40}$Ca + $^{40}$Ca collisions at 60$A$ MeV in the
centrality of 40\%--60\%. It is found that the hard-photon production
rate is sensitive to the density oscillations during the whole
reaction evolution. With the increase  in  density when the
collision system is in the compression stage, the system produces
more hard photons. In contrast, when the system starts to expand, the hard
photon production rate decreases. Actually, the density
oscillation of the colliding heavy-ion systems can be observed in
the experiments via hard-photon interferometry measurements
\cite{marques,schutz1,schutz} as well as dynamical dipole $\gamma$
radiation \cite{dipole,dipole2,Wu}. Apparently, hard photons are
mostly produced at the early stage of the reaction. Thereafter we
call these photons, emitted before the time of the first maximum
expansion of the system ($t \sim 65$ fm/c in Ca + Ca at 60$A$ MeV), as direct
hard photons (on the left side of the blue dashed line in
Fig.~\ref{time}(a)). It is also coincident with the definition in the  Sec.I of 
direct hard photons. We call the residual hard
photons producing in the later stage as thermal hard photons (on
the right side of blue dashed line in Fig.~\ref{time}). In this
way,  we can identify the producing photon as either direct or
thermal  by the emission time of photons in the present simulation.
We notice that the production rate of thermal photon tends to decrease for later time,
after 300 fm/c for Ca + Ca at 60$A$ MeV, which can be understood by the decrease of
 n-p collisions in an expanding system. However, the hard photon yields in the later stage after 300 fm/c
 is relative small fraction of the total thermal hard photon emission. Also, in the following calculations, we mainly focus
 on direct hard photons which  dominte the yield and do not depend on the  time evolution
 for the later stage.

\begin{figure}
 \includegraphics[scale=0.4]{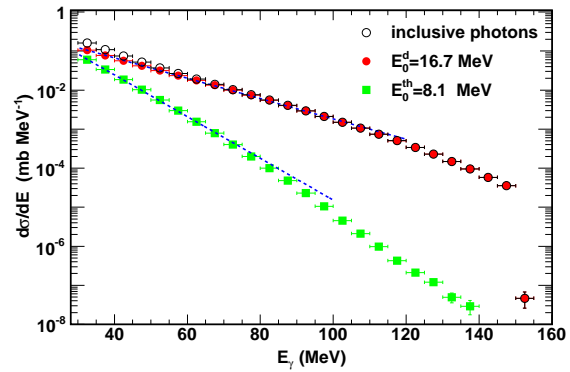}
\caption{\footnotesize  (Color online) Kinetic energy spectrum of
direct photons and thermal photons for inclusive events of
$^{40}$Ca + $^{40}$Ca at 60$A$ MeV with  the compressibility $K$ of 235 MeV.  Black dots and blue dots
represent the direct photons and thermal photons, respectively.
Lines represent the exponential fits with th Eq.~\ref{fit}. }
 \label{photon-spectra}
\end{figure}

For an example, Fig.~\ref{photon-spectra} demonstrates that the direct and thermal
hard photons for 60$A$ MeV $^{40}$Ca + $^{40}$Ca collisions with  a
compressibility $K$ of 235 exhibit different spectuml shapes: thermal photons give
rise to a softer energy spectral  than direct ones. In the
nucleon-nucleon cente-of-mass system, the photon spectrum can be
described  by the function:
\begin{equation}
 \frac{d\sigma}{d E_\gamma} = I_0 e^{-\frac{E_\gamma}{E_0}},
\end{equation}
where $I_0$ is a normalization constant and $E_0$ is the slope
parameter, which reflects the apparent source temperature of the 
photon emission source. The slope parameter $E_0$ depends on the
bombarding energy and on the average intrinsic momentum of the
participant nucleons. As less energy is available on average in
second-chance n-p collisions than in first-chance collisions
because most of the projectile kinetic energy is damped, the
thermal-photon spectrum becomes much softer. If one adds the two sources
together in the photon spectrum,  one obtains an empirical photon spectrum
of the form  \cite{schutz,martinez,enterria}:
\begin{equation}
 \frac{d\sigma}{d E_\gamma} = K_d e^{- \frac{E_\gamma}{E_0^d}} + K_{th} e^{-\frac{E_\gamma}{E_0^{th}}},
 \label{fit}
\end{equation}
with the constant $K_{d,th}$ defined by:
\begin{equation}
I_{d,th} = K_{d,th} \int_{E_{30}}^\infty
e^{-\frac{E_\gamma}{E_0^{d,th}} } dE_\gamma = K_{d,th} E_0^{d,th}
e^{-\frac{E_{30}}{E_0^{d,th}}}.
\end{equation}
$I_{d,th}$ represents the intensity of each photon source, $d$
stands for the direct photon and $th$ for stands for the thermal photon.
However, we notice that the change in slope
of the photon yield could be also affected by the 1/$E_{\gamma}$ factor
which enters the elementary ${np-np\gamma}$ bremsstrahlung probability \cite{Bona2_PR}.
 In this case, Eq.~\ref{fit} should be changed into
\begin{equation}
 \frac{d\sigma}{d E_\gamma} =  \frac{(K_d)'}{E_\gamma}  e^{- \frac{E_\gamma}{(E_0^d)'}} + \frac{(K_{th})'}{E_\gamma}  e^{-\frac{E_\gamma}{(E_0^{th})'}}.
 \label{fit2}
\end{equation}
In the next paragraph, we will compare the difference for the fitted slope with Eq.~\ref{fit} and Eq.~\ref{fit2}.

\begin{figure}
 \includegraphics[scale=0.4]{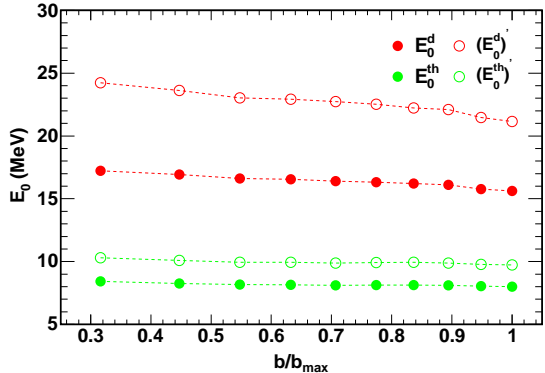}
\caption{\footnotesize  (Color online)  The slope parameters of
direct and thermal hard photons as a function of impact parameter
for $^{40}$Ca+$^{40}$Ca at 60$A$ MeV with the incompressibility $K$ of 235 MeV. Open symbols represent the results from the Eq.~\ref{fit2} and solid symbols are from Eq.~\ref{fit}. }
 \label{slope-b}
\end{figure}

As shown in the inset of the figure, the slope parametesr of direct
photons and thermal photons are 16.7 and  8.1 MeV,
respectively, for inclusive events  of $^{40}$Ca + $^{40}$Ca collisions  at
60$A$ MeV  with an incompressibility $K$ of 235 MeV if we fit the spectrum with  the Eq.~\ref{fit}. However, Eq.~\ref{fit2} will give
a larger slope value, namely 22.5 and 9.8 MeV, respectively, which is 35\% and 20\% higher than the fits without the 1/$E_{\gamma}$ factor (not plotted in the figure).
With two different fit formulas, we put the impact parameter dependence of the slope parameter together  in Fig.~\ref{slope-b}.
A remarkable difference is seen between the slope values with  and without  the 1/$E_{\gamma}$ factor, i.e.
 Eq.~\ref{fit2} gives a higher apparent photon temperature  than Eq.~\ref{fit}.
 In some previous experimental analysis \cite{schutz,martinez,enterria},   Eq.~\ref{fit} was mostly used to extract the apparent slope,
but this is probably not correct since a factor 1/$E_{\gamma}$ in front of the
exponential form can be derived from their elementary cross section [see Eq.~\ref{ddcs}].
On the other hand, the slope of hard photons displays
a slight decreasing behavior with  increasing  impact
parameter, indicating that the emission source becomes cooler
in peripheral collisions. Moreover, direct photons are  hotter
than thermal photons as we expected.

\subsection{Definition of anisotropic flows}

It is well known that collective flow is an important observable
in heavy ion collisions and it can carry some essential
information on the nuclear matter, such as the nuclear equation of
state
\cite{Olli,Voloshin,Sorge,Danile,Ma,Ma95,Zheng,Gale,INDRA,Yan,Chen}.
Anisotropic flows are  defined as  different $n$th harmonic
coefficients $v_n$ of an azimuthal Fourier expansion of the
particle invariant distribution \cite{Voloshin}
\begin{equation}
\frac{dN}{d\phi} \propto {1 + 2\sum_{n=1}^\infty v_n cos(n\phi) },
\end{equation}
where $\phi$ is the azimuthal angle between the transverse
momentum of the particle and the reaction plane. Note that in the
coordinate system the $z$-axis is along the beam axis, and the impact
parameter axis is labeled as the $x$-axis.

The first harmonic coefficient $v_1$ represents the directed flow,
\begin{equation}
v_1 = \langle cos\phi \rangle = \langle \frac{p_x}{p_t} \rangle,
\end{equation}
where $p_t = \sqrt{p_x^2+p_y^2}$ is the transverse momentum. The
second harmonic coefficient $v_2$ represents the elliptic flow
which characterizes the eccentricity of the particle distribution
in momentum space,
\begin{equation}
v_2 = \langle cos(2\phi) \rangle = \langle
\frac{p^2_x-p^2_y}{p^2_t} \rangle .
\end{equation}

\subsection{Directed  and elliptic flows of direct photons and free protons}

In relativistic heavy-ion collisions directed and elliptic flows of
hard photons have been recently reported in the experiments and through 
theoretical calculations \cite{Aggarwal,Adams,Turbide,Barz,Marques_96},
demonstrating a very useful tool for  exploring the properties of hot dense
matter. However, there are no experimental data available  so far  on
the flow of hard photons in intermediate energy heavy ion
collisions.
\begin{figure}[htbp]\centering
  \vspace{-0.1truein}
  \includegraphics[scale=0.43]{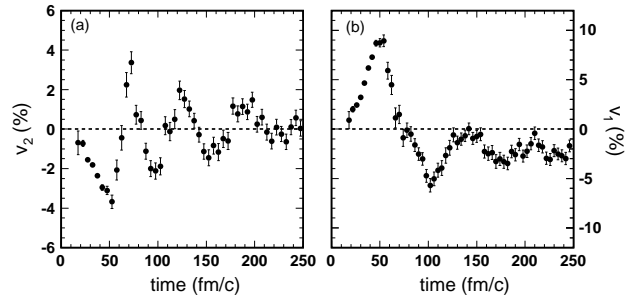}
  \vspace{0.1truein}
  \caption{\footnotesize The time evolution of elliptic flow ($v_2$) of hard photons (a) and
  directed flow ($v_1$) (b) for  60$A$ MeV Ca + Ca collisions at centrality of 40-60$\%$.
  }\label{flow_time}
\end{figure}
In this context, here we calculate directed transverse or elliptic
flows in intermediate energy heavy ion collisions. In addition,
given that hard photons mostly originate from bremsstrahlung
by individual proton-neutron collision and that free nucleons are also
emitted from nucleon-nucleon collisions, therefore it will be
interesting to compare  flows of protons and photons.

Fig.~\ref{flow_time} shows the time evolution of the average
values of directed flow and elliptic flow of the photons.
Considering the nearly symmetric behavior for directed flow versus
rapidity, here we calculate the average $v_1$ over only the
positive rapidity range, which can be taken as a measure of the
directed flow.

From Fig. 4, the directed flow of photons show rapid
rising with positive values during the compression stage and later
on it decreases to even negative value. Afterwards, the directed
flow remains negative since the system is never compressed. For
elliptic flow, the behavior shows contrary trend to the $v_1$ and
later on it shows oscillation for thermal photon emission. The
times corresponding to the peak or valley values of flows roughly
keep synchronized with the compression or expansion oscillation on
the system evolution as shown in Fig. 1(b).

\begin{figure}
  \vspace{-0.1truein}
   \includegraphics[scale=0.43]{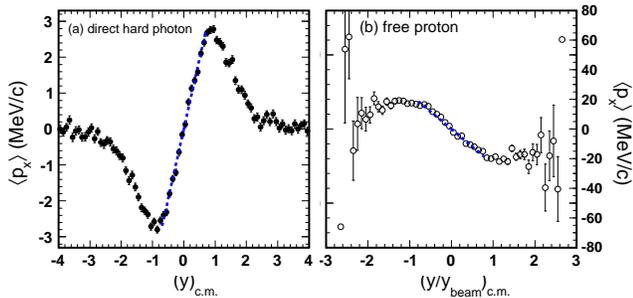}
\vspace{0.1truein}
  \caption{\footnotesize (Color online)   (a) Average in-plane transverse momentum of direct hard photons as a function of $c.m.$ rapidity
for 60$A$ MeV Ca + Ca collisions in the semi-central centrality of
$40-60\%$.
   The dashed line segment is a fit over the mid-rapidity region $-0.5 \leq y_{c.m.}\leq 0.5$.
   (b) Same as the left panel but for free protons.
   The dashed line segment is a fit over the mid-rapidity region $-0.5 \leq(y/y_{beam})_{c.m.}\leq 0.5$.
   }\label{flow}
\end{figure}

\begin{figure}
\includegraphics[scale=0.43]{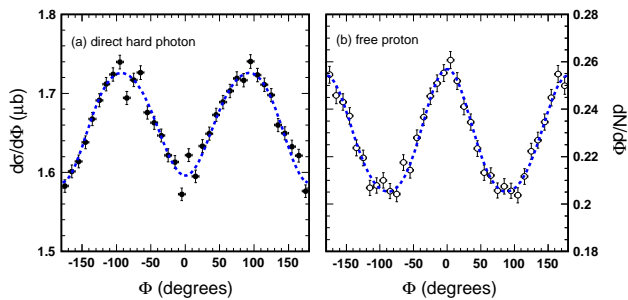}
  \caption{\footnotesize (Color online)  (a) and (b) is the azimuthal distribution of direct photons and
  of free protons, respectively, for 60$A$ MeV Ca + Ca collisions in
the semi-central centrality of  $40-60\%$. Both of them are fitted
to the 4-th order Fourier expansion. }\label{azimuthal}
\end{figure}

From the above calculations, we learn that  direct photons are
preferentially emitted in the out-of-plane (negative $v_2$) direction and
 thermal photons are emitted from a  thermalizing system which
makes their emission more isotropic (i.e. the oscillated
elliptic flow) than  the direct ones produced in  the pre-equilibrium
stage. In addition, thermal photons  contribute less than $30\%$
of the total yield of hard photon in the present model. Therefor
we will focus on  direct hard photons to discuss the flow results.

The quantity of directed transverse flow at mid-rapidity can be
also  defined by the slope:
  $F = \left. \frac{d \langle p_x \rangle}{d(y)_{c.m.}}
  \right|_{(y)_{c.m.}=0}$,
where $(y)_{c.m.}$ is the rapidity of particles in the center of
mass and $\langle p_x \rangle$ is the mean in-plane transverse
momentum of photons or protons in a given rapidity region. In
Fig.~\ref{flow}(a) and (b), we show  $\langle p_x \rangle$ plotted
versus the \textit{c.m.} rapidity $y_{c.m.}$ for direct hard
photons (a) as well as $\langle p_x \rangle$ plotted versus the
reduced \textit{c.m.} rapidity $(y/y_{beam})_{c.m.}$ for free
protons (b) for 60$A$ MeV Ca + Ca collisions in the semi-central
centrality of $40-60\%$. The EOS parameters with the
compressibility $K$ of 235  MeV are used.  The errors shown are
only statistical. For emitted proton (free proton), we identify it
in the BUU calculation as its local density $\rho < 1/8 \rho_0$.
We take the values of flows when the system has been already in
the freeze-out time at 180 fm/c. A good linearity is seen in the
mid-rapidity region $(-0.5, 0.5)$ and the slope of a linear fit
can be  defined as the magnitude of directed transverse flow. The
extracted value for the directed transverse flow of direct hard
photons is about $+3.9$ MeV/c, and that of free protons is about
$-23.7$ MeV/c. Thus direct hard photons do exist the directed
transverse flow even though the absolute value is smaller than the
proton's flow, and the sign of its flow is just opposite to that
of free protons.

\begin{figure}
\vspace{-0.1truein}
     \includegraphics[scale=0.43]{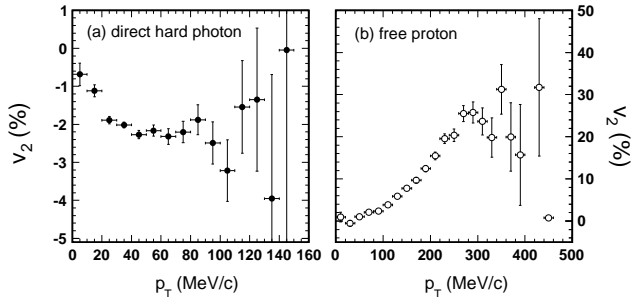}
 \vspace{0.1truein}
  \caption{\footnotesize   Elliptic flows ($v_2$) for direct hard photons (a) and free protons
(b)   as a function of transverse momentum ($p_{T}$) for 60$A$ MeV
Ca + Ca collisions in the semi-central centrality of $40-60\%$.
  }\label{v2Pt}
\end{figure}

In order to extract the value of elliptic flow and reduce the
error of fits, we fit the azimuthal distribution to  the $4th$
order Fourier expansion. Fig.\ref{azimuthal} shows an example for
such fits with the elliptic flow parameter $v_2$= -2$\%$ for
photons and  $v_2$= 5.5 $\%$ for protons.

Fig.~\ref{v2Pt} shows the differential elliptic flow of direct
hard photons (a) and of free protons (b) as a function of
transverse momentum $p_{T}$ for $^{40}$Ca + $^{40}$Ca at 60$A$
MeV. Similar to the directed transverse flow, the values of
elliptic flow of direct hard photons and of free protons also have
an opposite signs at this reaction energy, i.e., reflecting a
different preferential transverse emission in the direction of
out-of-plane or in-plane, respectively. Meanwhile, absolute
flow values for photons are smaller than the proton's  like
the behavior of transverse flow. Except of the opposite sign, we see
that both elliptic flows have similar tendency with the increasing
of $p_{T}$, i.e., their absolute values increase at lower $p_{T}$,
and becomes gradually saturated, especially for direct hard
photons.

To explain the correlation of the collective flow between direct
hard photons and free protons indicating above, we should note
that flows of direct hard photons originate from the individual
proton-neutron collisions. As Eq.~\ref{ddcs} shows, we can roughly
consider that in the individual proton-neutron center of mass
system, in directions perpendicular to incident proton velocity,
i.e. $\theta_{\gamma}=90^{\circ}$, the possibility of hard photon
production is much larger than the parallel direction, actually it
is also in agreement with the theoretical calculations and the
experiments \cite{Herrmann,Safkan}. As a whole, the flow of hard
photons should be correlated with the collective movement of the
nucleons, and presents an opposite behavior. Consequently, flows
of hard photons and of free nucleons exhibit an opposite
character.

\section{ Dependences of hard photon production and flow on some variables}

\subsection{Impact parameter dependence}

It is well known that anisotropic flow mainly originates from the
initial asymmetric overlap zone of colliding nuclei which induces
different pressures  or rotational collective motion of the
participant region and leads to anisotropic emission of the
particles. Peripheral collisions have more initial asymmetry in
overlap zone than  central collisions, thus  more anisotropic
emission of the particles is expected. Therefore anisotropic
emission shall be sensitive to impact parameter.

\begin{figure}
\begin{center}
\vspace{+0.1truein}
     \includegraphics[scale=0.43]{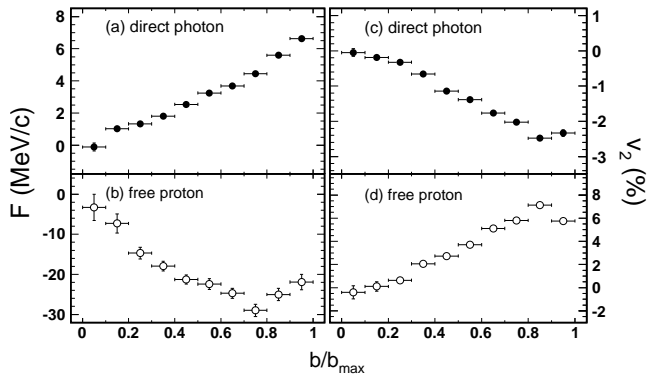}
\caption{\label{fig8}  (a) and (b) are the directed transverse
flow parameter $F$ of direct photons and free protons,
respectively, as a function of the reduced impact parameter for
the reaction
 $^{40}$Ca + $^{40}$Ca  collisions at 60$A$ MeV. (c) and (d) same
as above but for elliptic asymmetry coefficient $v_2$.}
\end{center}
\end{figure}

Here we simulate the reaction  $^{40}$Ca + $^{40}$Ca  collisions at 60$A$ MeV, and use the
EOS parameters with the compressibility $K$ of $235$ MeV for the nuclear mean
field $U$. Fig.~\ref{fig8} (a) and (b) shows the directed
transverse flow parameter $F$ of direct photons and free protons,
respectively, as a function of the reduced impact parameter (i.e.
normalized by the maximum impact parameter). Fig.~\ref{fig8} (c)
and (d) are the same as (a) and (b) but for elliptic asymmetry
coefficient $v_2$. We can see that both $F$ and $v_2$ of direct
photons and free protons have the similar tendency with impact parameter, i.e. their absolute values increase
with the impact parameter except of slightly decreasing in very
peripheral collisions for $v_2$. We also see that at all impact
parameters, in contrast to  to the negative directed transverse flow and
positive elliptic flow of free protons, direct photons show the
positive $F$ and the negative $v_2$, i.e. the anisotropy is
shifted by a phase $\pi$/2. It agrees with the previous conclusion
that the azimuthal asymmetry of direct photons is anti-correlated
with the corresponding free proton's flow.

\subsection{Beam energy dependence}

\begin{figure}
\begin{center}
     \includegraphics[scale=0.43]{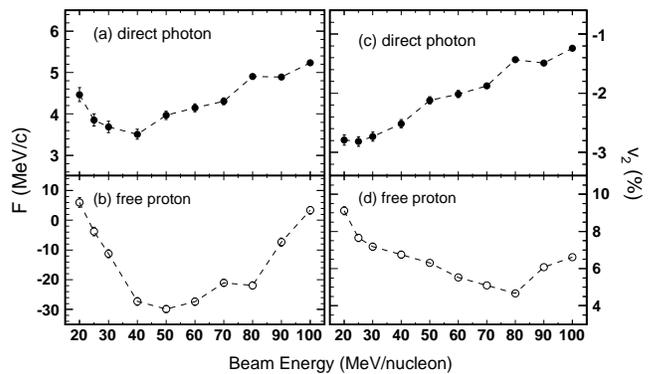}
\caption{  (a) and (b) is the directed transverse
flow parameter $F$ of direct photons and free protons,
respectively, as a function of beam energy for the reaction which
is the same as Fig.~\ref{fig8} but only semi-central events
$(40-60\%)$. (c) and (d) same as above but for elliptic asymmetry
coefficient $v_2$.} \label{v2-E}
\end{center}
\end{figure}

In Fig.~\ref{v2-E} we show the directed transverse flow parameter
$F$ and elliptic asymmetry coefficient $v_2$ of direct photons and
free protons, respectively, as a function of beam energy for the
Ca + Ca reaction at $(40-60\%)$ centrality. In the beam energy
range studied here, same as the impact parameter dependence, the
opposite signs of $F$ and $v_2$ between direct photons and free
protons  are also  anticorrelated. Moreover, except for the
opposite sign, directed transverse flow parameter $F$ of direct
photons and free protons have similar structures with the
increasing of beam energy. The value of direct photon elliptic
asymmetry coefficient $v_2$ increases with beam energy from
negative to positive, a tendency  similar to available
experimental results on hard photons in relativistic
heavy-ion collisions.

\subsection{EOS and symmetry energy dependences}

In this section we will discuss the nuclear equation of state
and its symmetry energy term depdences of  hard photon production. In our calculations, the
single-particle potential taken as the Skyrme parameterized mean
field potential including symmetry potential is shown below
\begin{equation}
U^{n(p)}(\rho) = \alpha (\frac{\rho}{\rho_0}) + \beta
(\frac{\rho}{\rho_0})^\sigma + V_{asy}^{n(p)} (\rho,\delta),
\end{equation}
where  the coefficients $a$, $b$ and $\sigma$ are parameters for
nuclear equation of state which is determined  by the nuclear
saturation property and the incompressibility $K$ of symmetric
nuclear matter; $\delta$ is the isospin-asymmetry parameter,
$\delta = (\rho_n-\rho_p)/(\rho_n+\rho_p)$, and $V_{asy}^{n(p)}
(\rho,\delta)$ is the symmetry potential of neutrons (protons).

First, we only investigate the effects of the incompressibility
$K$ on the hard photon production in symmetric reaction system. In
the Skyrme potential, we take the  incompressibility $K$  as 200
MeV, 235 MeV and 380 MeV, in which the first two correspond to
soft and semi-soft potential and the last one corresponds to hard
potential as we introduced in Sec.II A.

\begin{figure}
\includegraphics[scale=0.4]{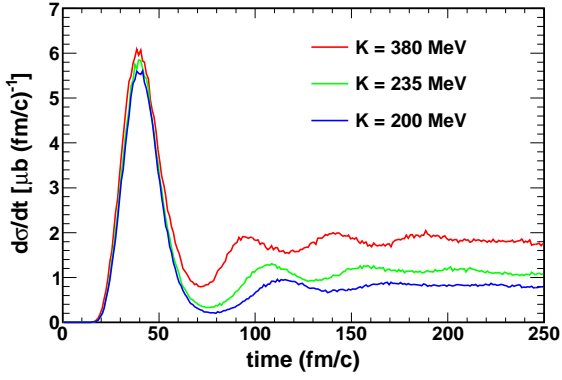}
 \caption{\footnotesize  (Color online) Time evolution of production rate of
 bremsstrahlung photons with time for inclusive events of $^{40}$Ca + $^{40}$Ca at 30$A$ MeV.
 Different EOS are used, namely the hard EOS, the semi-soft EOS and the soft EOS, respectively.}
 \label{product-rate}.
 \end{figure}

Because of the sensitivity to the density oscillations of
colliding system, hard photon should be also dependent of the EOS
of nuclear matter, especially for thermal hard photons
\cite{schutz1,schutz2}. Actually as shown in
Fig.~\ref{product-rate}, photon production shows its sensitivity
to the compressibility, especially for thermal photon which are
produced after $t \sim 80$ fm/c. It shows that the stiffer the
EOS, the higher multiplicity the thermal photons. In contrast, 
direct photons are  produced by the first channel neutron-proton
bremsstrahlung, their production rate only weakly depends on EOS
since direct hard photons are emitted by the first channel n-p
scattering when the system is in a highly nonequilibrium state
during the compression stage,  and they don not have enough times to feel
the EOS.

\begin{figure}[htbp]
\includegraphics[scale=0.4]{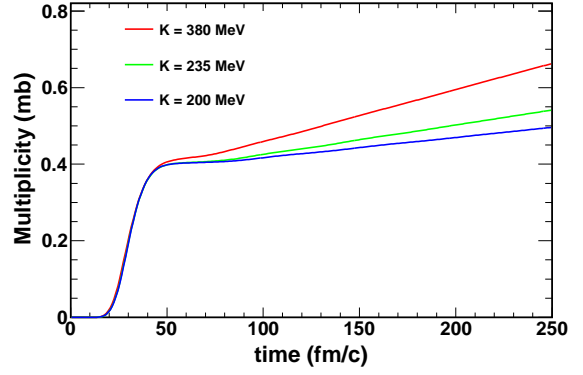}
\caption{(Color online) Time evolution of hard photon multiplicity
with different incompressibility $K$ for the reaction  $^{40}$Ca +
$^{40}$Ca  at 60$A$ MeV.} \label{fig:M_t_EOS}
\end{figure}

Shown in Fig.~\ref{fig:M_t_EOS} is the time evolution of inclusive
hard photon multiplicity from the symmetric reaction
system  $^{40}$Ca + $^{40}$Ca at 60$A$ MeV. We can see that at
this energy,  for different incompressibility $K$, multiplicities
of hard photons produced in the early stage of collisions are
nearly equivalent and later they show a clear correlation with
incompressibility $K$. Corresponding to direct and thermal
photons, this indicates that direct photon production is not
sensitive to  incompressibility $K$, because at this reaction
energy, comparable with two-body interactions, mean field can be
neglected in producing hard photons in the early stage of
collisions. However, there is a  correlation  between $K$ and thermal
photon production, with larger $K$ producing more thermal photons.

\begin{figure}
\includegraphics[scale=0.4]{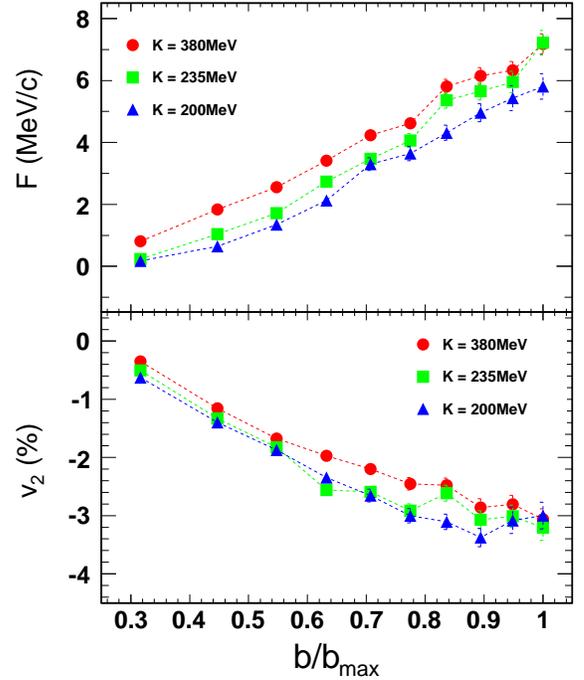}
\caption{\footnotesize  (Color online) Directed flow parameter
($F$) and elliptic flow ($v_2$) as a function of impact parameter
with different EOS parameters for Ca + Ca at 30$A$ MeV.  }
\label{flow-EOS}
\end{figure}

For the magnitude of flow parameters $F$ and $v_2$, comparisons  are made for  different
equation of state. Fig.~\ref{flow-EOS} shows the
$F$ and $v_2$ as a function of impact parameter with different
EOS. Generally, the directed flow becomes larger with the
increasing of stiffness of the EOS, whereas the elliptic flow becomes smaller.

In order to study the effects of symmetry potential, we now set
incompressibility $K$ as 380 MeV. We mainly consider three kinds
of symmetry potential: (1)  the symmetry potential is neglected,
that is
\begin{equation}
V_{asy}^{(0)}(\rho,\delta) = 0; \end{equation}
(2) the symmetry
potential is a linear function of isospin-asymmetry parameter
$\delta$, that is
\begin{equation}
V_{asy}^{(1)}(\rho,\delta)=C_{sym}\delta\tau^{n(p)},
\end{equation}
 where
$C_{sym}$=$32$ MeV, $\tau^n$=1, $\tau^p$=-1; \\
(3) the Li et al. single-particle potential  from the parametrization of nuclear symmetry energy used in
Ref.~\cite{Heiselberg} for studying the properties of neutron
stars,  that is
$E_{sym}(\rho)=E_0(\rho_0)(\frac{\rho}{\rho_0})^{\gamma}$,  Li
{\it et al.} derived a single-particle symmetry potential
as~\cite{LBA}:
\begin{equation}
\begin{aligned}
V_{asy}^{(2)}(\rho,\delta)=&\left[E_0(\rho_0)(\gamma-1)(\frac{\rho}{\rho_0})^{\gamma}+4.2(\frac{\rho}{\rho_0})^{2/3}\right]\times
 \delta^2\\
 &+\left(E_0(\rho_0)(\frac{\rho}{\rho_0})^{\gamma}-12.7(\frac{\rho}{\rho_0})^{2/3}\right)\delta\tau^{n(p)},
% \nonumber
\label{VasyLBA}
\end{aligned}
\end{equation}
where $\tau^n$=1, $\tau^p$=-1, $E_0(\rho_0)$ sets as 30 MeV in
this paper,  and parameter $\gamma$ represents the stiffness of the
symmetry energy. In the following, we consider two cases of
$\gamma$ = 0.5  and 2, respectively, corresponding to soft and
stiff symmetry energy, to explore the effects of different
symmetry potential.

\begin{figure}
\includegraphics[scale=0.4]{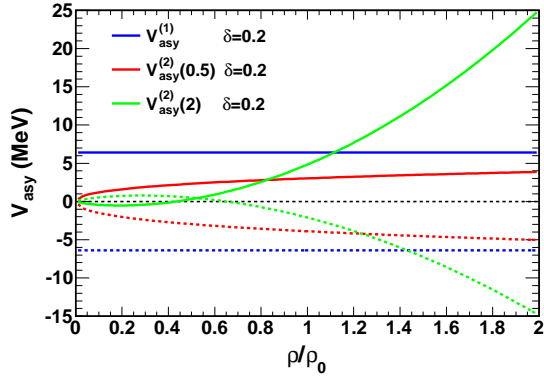}
\caption{\footnotesize  (Color online)  Different symmetry
potential which is used in the present model calculation as a
function of density for an isospin asymmetry of $\delta$ = 0.2 and
the $\gamma$ parameters of 0.5 (red line) and 2.0 (green line),
respectively. Solid lines are for neutrons and dashed lines for
protons. }\label{potential}
\end{figure}

The symmetry potentials for neutron and proton with $\delta$ =
0.20 is plotted in Fig.~\ref{potential}, where the red line
represents $V_{asy}^{(1)}$, and the blue and green lines represent
$V_{asy}^{(2)}$ for  $\gamma$ parameters of 0.5 and 2.0 for
neutrons (upper branch) and protons (lower branch), respectively.
For small isospin asymmetry  and density  near $\rho_0$ the
above symmetry potentials reduce to the well known Lane potential
which varies linearly with $\delta$ \cite{Lane}.  Generally, the
repulsive/attractive symmetry potential for neutrons/ protons
increases with density.

\begin{figure}
\includegraphics[scale=0.4]{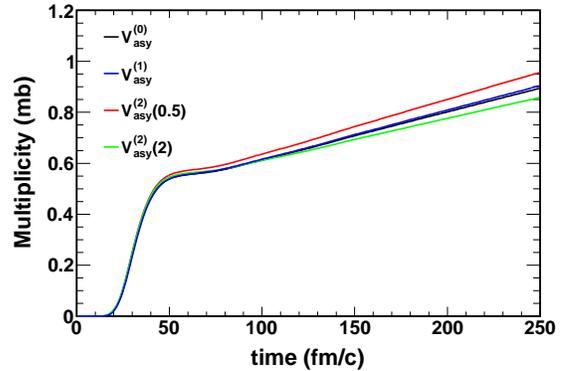}
\caption{\label{fig:VasyCa48LBA} (Color online) Time evolution of
hard photon multiplicity with different symmetry potentials for
the reaction $^{48}$Ca + $^{48}$Ca at 60$A$ MeV.}
\end{figure}

Fig.~\ref{fig:VasyCa48LBA} shows the time evolution of inclusive
hard photon multiplicity in the reaction $^{48}$Ca + $^{48}$Ca at
60$A$ MeV. In order to separate the effects of symmetry potential
from the incompressibility, we set $K$ as a constant 380 MeV.
Fig.~\ref{fig:VasyCa48LBA} presents the effects of different
symmetry potentials on hard photon productions. Same as the
incompressibility $K$, direct photon production is insensitive to
symmetry potential at the studied reaction energy. For thermal photon, the
curves of $V_{asy}^{(0)}$ and $V_{asy}^{(1)}$ have the nearly same
multiplicities with time evolution, and they are intermediate
between $V_{asy}^{(2)}(0.5)$ and $V_{asy}^{(2)}(2)$. We note that
thermal photon production is sensitive to  $\gamma$-parameter in
symmetry potential $V_{asy}^{(2)}(\gamma)$,  i.e. less $\gamma$,  corresponds to softer symmetry
energy, induces more thermal photons. And we also note that
the sensitivity of thermal photon production to $\gamma$ is not  remarkable
in comparison with its sensitivity to the incompressibility $K$, which is of course
understandable since the symmetry energy is relative small term in
comparison with the incompressibility $K$.

From the effects of the incompressibility and symmetry potential
on hard photon production, it appears that the yield of thermal
photons has rather strong dependence on nuclear compressibility as
well as symmetry energy, whereas the yield of direct
photons is rather insensitive to nuclear EOS. Therefore thermal hard photon could serve
as a porble of nuclear EOS in intermediate energy HIC.

\section{Photon-photon correlations}

In above sections, we mainly focus on the properties of inclusive hard photons, from which
we find that production distribution and anistropic flows of hard photons reveal rich information of nuclear
dynamics. Also, the sensitivity to the nuclear equation of state and symmetry energy for
hard photons has been discussed.  In this section, we will discuss the correlation properties
of photons in terms of two-particle correlation technique that can provide us a very powerful tool
to characterise the properties of a particle source. In particular, two-boson relative momentum distributions
enable one to determine the space-time structure according to the formalism of Bose-Einstein correlations.
The magnitude of the correlation can be related to ther space-time distribution of the boson source.
In the follwoing calculations, we will construct two-photon momentum correlation function as well as
 azimuthal correlation function,
from which the space-time structure of the photon souce  and anistropy property can be indicated.

\subsection{Two-photon momentum correlation}

Intensity interferometry (also called HBT correlation) is  used as a universal tool to study
the properties of any boson sources such as stars \cite{HBT}, or
photon and meson sources in the early phase of heavy-ion
collisions \cite{Goldhaber}. The formalism was developed starting
from optics and quantum statistics and was finally adapted to the
dynamics of heavy-ion collisions
\cite{Lorstad,Boal,Quebert,Koonin,Pratt,Lynch,Marques_RP,Ma-hbt,Ma-hbt2}.
We have performed the calculation which evaluates the correlation function directly from the photon source distribution given by the BUU calculation.
We store for each $i$th $pn$ collision its position
$\overrightarrow{r_{i}}$ and the associated photon probability
distribution $P_{i}(\overrightarrow{k_{i}})$. After the completion
of the calculation, we analyze this output data to construct plane
waves with four-momentum $\overrightarrow{k_i}$ at
$\overrightarrow{r_i}$ and calculate the two-photon probability
for $i \neq j$ as \cite{Barz,Marques2}:
\begin{eqnarray}
P_{12} &= P_{1 \bigotimes 2} \left | A e^{i(\overrightarrow{k_{1}}
\cdot \overrightarrow{r_{i}} + \overrightarrow{k_{2}} \cdot
\overrightarrow{r_{j}})} + B
e^{i(\overrightarrow{k_{1}} \cdot \overrightarrow{r_{j}} + \overrightarrow{ k_{2}} \cdot \overrightarrow{r_{i}})} \right |^{2} \\
& = P_{1 \bigotimes 2} \left [ 1 + 2AB\cos[
q(\overrightarrow{r_{i}}-\overrightarrow{r_{j}})]\right],
\label{two_photon_pro}
\end{eqnarray}
where $P_{1 \bigotimes 2}$ represents the probability to produce a
pair without correlation, and $A$ and $B$ are the amplitudes
related to the normalized probabilities of the direct $\left [
P_{i}(\overrightarrow{k_{1}})P_{j}(\overrightarrow{k_{2}}) \right
]$ and cross terms $\left
[P_{j}(\overrightarrow{k_{1}})P_{i}(\overrightarrow{k_{2}})
\right]$. This corresponds to Fourier transforming the photon
source event by event. We set the weight of the interference in
Eq.~\ref{two_photon_pro} as $\kappa$, that is $\kappa = 2AB$,
which was predicted between 0.5 and 1.0 depending on the
anisotropy of hard photon emission. The exact experimental filter
was finally applied to the projection onto the Lorentz-invariant
relative four momentum $Q = \sqrt{\mathbf{q}^{2}-q_{0}^{2}}$ of
the resulting distribution $P_{12}$ and $P_{1 \bigotimes 2}$ of
Eq.~\ref{two_photon_pro}, then the two-photon correlation function
was calculated as $C_{12}(Q) = \frac{P_{12}}{P_{1 \bigotimes 2}} =
1+ \kappa\cos[q(\overrightarrow{r_{i}} - \overrightarrow{r_{j}})]
= f(Q)$.

For convenience, $\kappa$ was set to 0.75 for our calculations in
order to take into account the established
anisotropic component in the angular distribution of hard photons.
   The two-photon correlation function has been fitted with a
Gaussian parameterization with correlation strength $\lambda$ and
radius parameter $R_{Q}$:
\begin{equation}
C_{12} = 1 + \lambda exp(-Q^{2}R_{Q}^{2}). \label{hC_fit}
\end{equation}
In the above relation $R_{Q}$ is the space-time parameter
conjugate to $Q$, which measures an invariant
length depending
on the source-size parameters $R$ and $\tau$:
\begin{equation}
R_{Q}=
R\sqrt{\frac{1+(\tau/R)^{2}(q_{0}/q)^{2}}{1-(q_{0}/q)^{2}}}.
\end{equation}
One can easily know that, since $q_{0} \ll q$, $R_{Q}$ is a first
order measure of the spatial extent of the source, that is $R
\approx R_{Q} $. We can then calculate from $R_Q$ the rms radius
of the source as the one of a static three-dimensional Gaussian
source: $R_{rms}=\sqrt{3}R_Q$.

\begin{figure}[htbp]
\vspace{-0.1truein}\centering
 \includegraphics[scale=0.4]{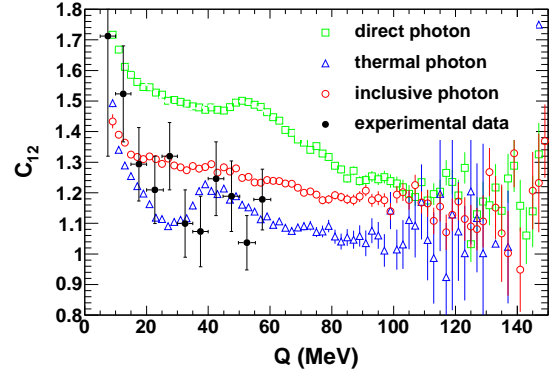}
  \caption{\footnotesize  (Color online) Two-photon HBT correlation function
  for the reaction  $^{86}$Kr + $^{58}$Ni collisions at 60$A$ MeV in the laboratory frame, $E_{\gamma}>25$ MeV. The green
squares represent
  for direct photons, blue triangles for thermal photons, and red
  open circles for inclusive photons. The black closed circles
  represent for the experimental data taken from Ref.~\cite{Barz,Marques_96}.
  }\label{C_hbt}
\end{figure}

\begin{figure}
\vspace{-0.1truein}
 \includegraphics[scale=0.4]{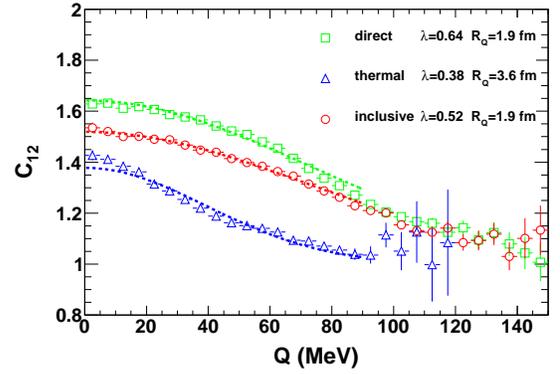}
  \caption{\footnotesize  (Color online) Two-photon HBT correlation function
  for the reaction  $^{40}$Ca + $^{40}$Ca  collisions at 60$A$ MeV. The symbols are the same as Fig.~\ref{C_hbt} and the
  dashed lines are fitting functions as Eq.~\ref{hC_fit}.
  }\label{C_hbt2}
\end{figure}

For the comparisons with experimental data, we calculated  the
reaction $^{86}$Kr + $^{58}$Ni at 60$A$ MeV, employing here the
following filter which is very similar to the experiment:
$E_\gamma(1,2) > 25$ MeV, detector positions between polar angles
of $35^\circ$ and $165^\circ$ (orienting downstream), azimuthal
opening angles of $0\pm28^\circ$ and $180\pm28^\circ$, and
$18^\circ$ for the minimum opening angle. Fig.~\ref{C_hbt}
presents HBT correlation functions of direct photons (green
squres), thermal photons (blue triangles) and inclusive photons
(red open circles), respectively. We find that the correlation
function of direct photons is much larger than thermal one, and
the correlation function of inclusive photons is intermediate.
This result is reasonable, because direct photons are emitted at
the early compressed stage of collisions, so they have stronger
interference than thermal photons which are produced in later
thermalizing stage. We also see that our result can well
reproduce the experimental data \cite{Barz,Marques_96}, especially for the correlation
function of thermal photons, which agrees with the experimental
function even the oscillation structure. Therefore, we can successfully
reproduce the HBT correlation function of hard photons by the BUU
simulation .

In order to further study the HBT correlation of hard photon and
extract photon source information, we calculated the
symmetric reaction system  $^{40}$Ca + $^{40}$Ca at 60$A$ MeV in
the C.M. frame, and only the events of central collision
($0<b_{red}\leq 0.2$) were taken for simplification. In
Fig.~\ref{C_hbt2}, we find that the correlation function of
direct photons is the largest, the thermal one is the least, and the correlation function of 
inclusive photons are between them. Moreover, in the conditions of
above reaction, all of them can be well fitted by
Eq.~\ref{hC_fit}. After fitting correlation function, we can
obtain two useful fitting parameters: correlation strength
$\lambda$ and radius parameter $R_{Q}$. As the results on top
right corner show, direct photons have the largest correlation
strength $\lambda$, the second is inclusive photons and the least
one is thermal photons. Thus $\lambda$ is a parameter which is
sensitive to the intensity of interference. And we also get three
radius parameters $R_{Q}$. $R_{Q}$ of thermal photons is nearly
twice than direct photons', and inclusive photon is equivalent to
the later. To explain this, we know that direct photon is mostly
emitted in the early stage when the reaction system is strongly
compressed, so the emission source of photons is small. And later
on, the thermalizing system extends very much in company with
the production of thermal photons. In this reaction energy at
$E_{lab}$ = 60$A$ MeV, most of hard photons are produced in the
early stage of the collision, so the spatial source extent of
inclusive photons should be approximate to the direct photons. As
a result, they have equivalent radius parameter $R_{Q}$. Then we
can get the rms radius of the photon source by the equation
$R_{rms}=\sqrt{3}R_Q$. Therefore, two-photon correlation function
provides the information of hard photon source, which is available
to investigate the emission source during the collisions.

\subsection{Two-photon azimuthal correlation}

From individual-photon azimuthal distribution, we find that direct
hard photons exhibit the azimuthal asymmetric emission in
intermediate energy heavy-ion collisions, especially negative
elliptic flow parameter $v_{2}$. Actually it is difficult to
extract the elliptic flow parameter $v_{2}$ by the method of
reconstructing  reaction plane in the experiment, so as below,
we will discuss to use the method of two-photon azimuthal
correlation to extract the elliptic flow of direct photons
directly.

For particles in the same class, we  defined the particle
azimuthal correlation function following the multi-fragment
azimuthal correlation method \cite{Ma95,Wang,Lacey}. The multi-fragment
azimuthal correlation function is defined as follows:
\begin{equation}
C(\Delta \phi)=\frac{N_{cor}(\Delta \phi)}{N_{uncor}(\Delta \phi)},
\label{corfun}
\end{equation}
where $N_{cor}$ is the $\Delta \phi$ distribution for fragment
pairs from the same event and $N_{uncor}$ is the $\Delta \phi$
distribution by randomly selecting each member of a fragment pair
to form mixed events. The $\Delta \phi$ between all selected
fragments in an event are used to construct the correlation
function, that is, $\frac{n(n-1)}{2}\Delta \phi$ angles for $n$
fragments.

In our calculations for two-photon azimuthal correlations, we get
$N_{cor}$ from the $\Delta \phi$ distribution of photon pairs in
the same event and $N_{uncor}$ from the mixed events. For these
correlations, we may make a fit of the Fourier series with the
expression
\begin{equation}
C(\Delta\phi)=A[1+\lambda_{1}\cos(\Delta\phi)+\lambda_{2}\cos(2\Delta\phi)],
\label{Cazifit}
\end{equation}
where $\lambda_{1}$ and $\lambda_{2}$ are treated as fit
parameters.

\begin{figure}
\includegraphics[scale=0.4]{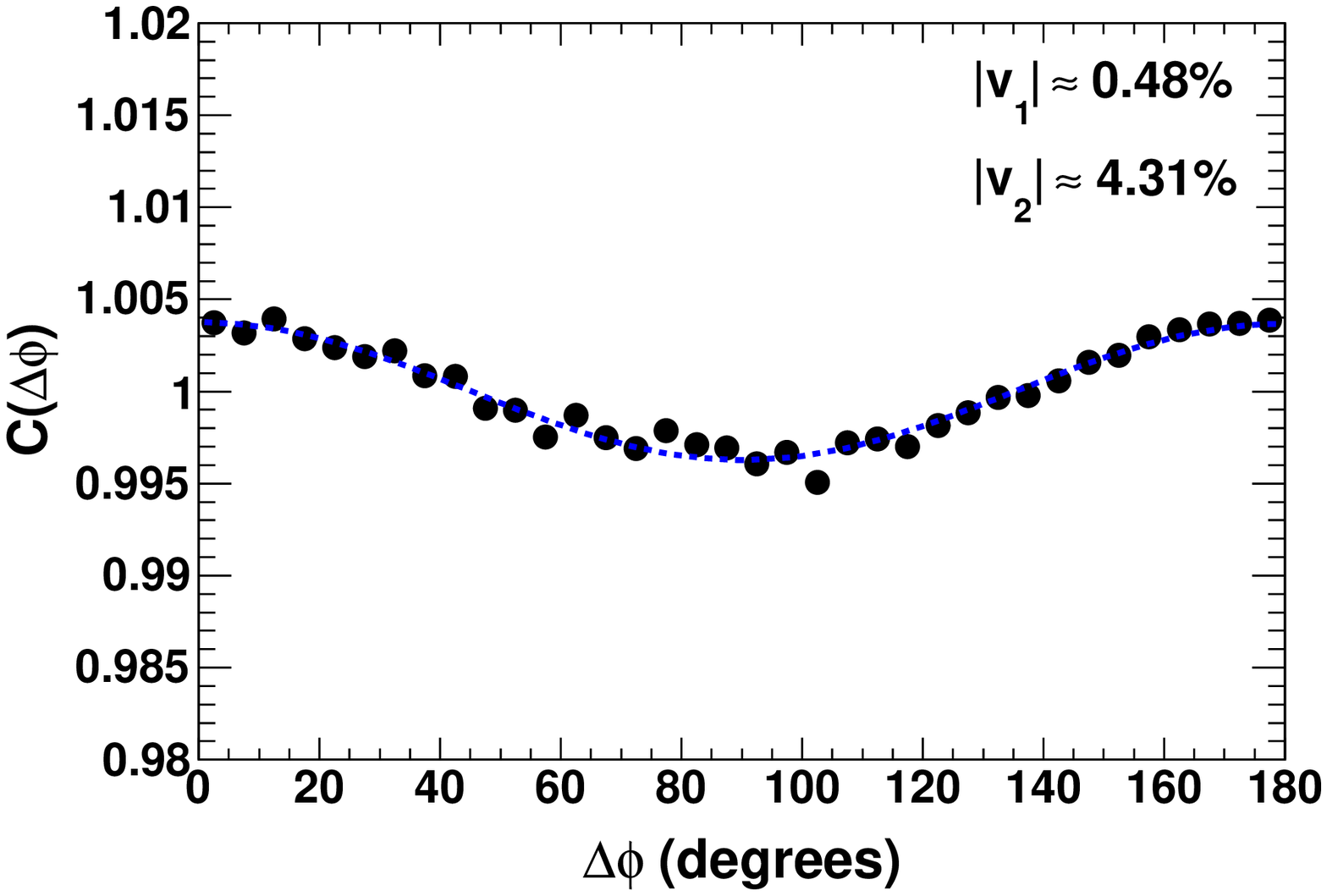}
\includegraphics[scale=0.4]{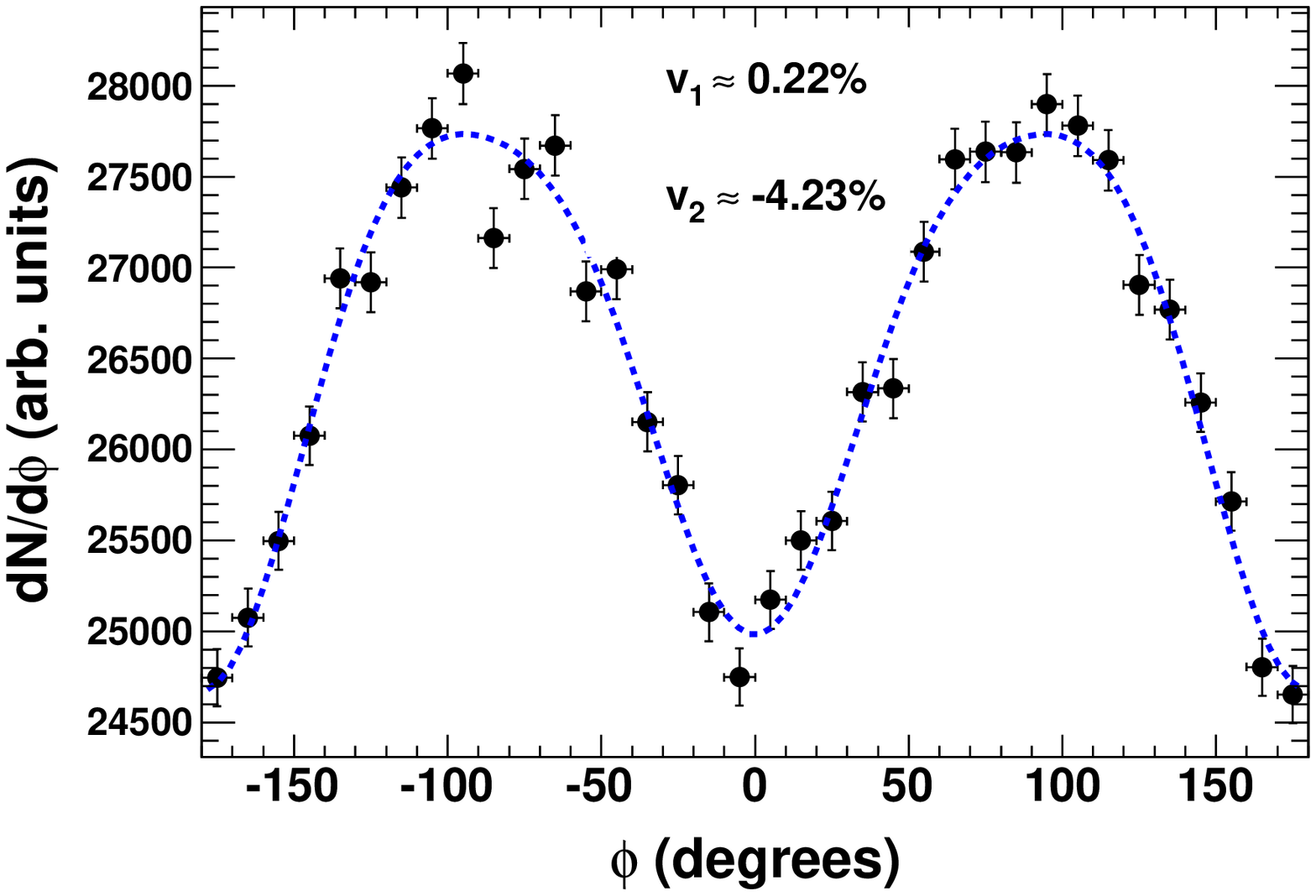}
  \caption{\footnotesize (Color online) Top panel: Two-photon azimuthal correlation function
  for the reaction   $^{40}$Ca + $^{40}$Ca  at 60$A$ MeV for semi-central events
  (40\%--60\%), and the dashed line is the fitting function as
  Eq.~\ref{Cazifit};
Bottom panel: Individual-photon azimuthal distribution  for the
same reaction, and the dashed line is a fit
  of Fourier expansion.}
\label{C_azi}
\end{figure}

Under the assumption of statistically independent emission of
particles with the same azimuthal distribution $F(\phi)$ in an
event, the
azimuthal correlation function is simply related to
$F(\phi)$ via the convolution
\begin{equation}
C(\Delta\phi)=\int_0^{2\pi} F(\phi)F(\phi+\Delta\phi)d\phi .
\label{Fourier}
\end{equation}

One the other hand, $F(\phi)$ can be described by the Fourier
expansion:
\begin{equation}
F(\phi)=\frac{dN}{d\phi} \propto 1 + 2 \sum_{n=1}^{\infty}
v_{n}\cos(n\phi), \label{azi_dist}
\end{equation}
where the different $n$th harmonic Fourier expansion coefficient
$v_{n}$ is defined as the $n$th anisotropic flow,  of which $v_{1}$ is directed flow
parameter and $v_{2}$ is so called elliptic flow parameter.

If we only take $v_1$ and $v_2$ terms in Eq.~\ref{azi_dist} and
substitute Eq. ~\ref{azi_dist} into Eq. ~\ref{Fourier}, we can
derive the form of $C(\Delta\phi)$ as follows:
\begin{equation}
C(\Delta\phi) = B [1 + 2 v_{1}^{2}\cos(\Delta\phi) + 2
v_{2}^{2}\cos(2\Delta\phi)]  \label{C_derive}
\end{equation}

From Eq.\ref{Fourier} and Eq.\ref{azi_dist}, we then get the
relationship between the fitted parameter $\lambda_{2}$ and the
elliptic flow parameter $v_{2}$~\cite{Ma95}:
\begin{equation}
|v_{2}|\approx\sqrt{\lambda_{2}/2}.
 \label{v2_lambda}
\end{equation}

\begin{figure}\centering
  \includegraphics[scale=0.43]{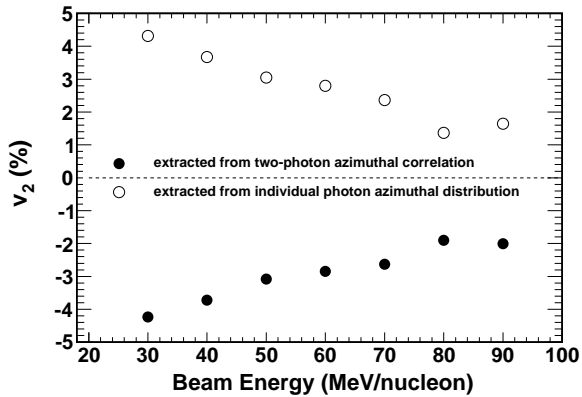}
 \vspace{0.1truein}
  \caption{\footnotesize   Excitation function of $v_2$
   for reaction  $^{40}$Ca + $^{40}$Ca  at 60$A$ MeV with centrality 40\%--60\%, open
   circles represent the amplitude of $v_2$ extracted from two-photon
   azimuthal correlation and red ones from the fit of Fourier
   expansion to individual-photon azimuthal distribution.
  }
\label{hCazi_v2E}
\end{figure}

Now we make the comparison of elliptic flow parameter $v_{2}$
extracted from two-photon azimuthal correlation with fitting value
from individual-photon azimuthal distribution. In
Fig.~\ref{C_azi}, we calculated the reaction   $^{40}$Ca +
$^{40}$Ca at 60$A$ MeV for semi-central events (40\%--60\%) with
the compressibility $K$ = 235 MeV, and we extracted the elliptic
flow parameters $v_{2}$ by the two distinct methods above. Then we
can see that the absolute values of $v_{2}$ are nearly equivalent,
except that we can only get the
amplitude of elliptic flow  from two-photon azimuthal correlation, and the fitting value of $v_2$ from
individual-photon azimuthal distribution is negative. Further, as
shown in Fig.~\ref{hCazi_v2E}, we get the excitation function of
$v_2$ by the above two methods  which shows nearly
same amplitude of $v_2$ in various reaction energy but different
signs. Therefore, in view of the difficulty to reconstruct
reaction plane in experiment, two-photon azimuthal correlation
provides us an alternative method to extract the amplitude of
elliptic flow parameter $v_{2}$ of hard photons.

\section{Summary}

In summary, we have systematically investigated hard photon
production by the process of proton-neutron bremsstrahlung as well
as the behavior of azimuthal asymmetry of hard photons and free
protons produced in intermediate energy heavy-ion collisions in
the framework of the BUU model. As discussed in previous studies,
hard photons can be separated into direct hard photon which is produced
by the first channel neutron-proton bremsstrahlung and thermal
hard photon which is produced by the later stage neutron-proton
bremsstrahlung during the system's evolution towards thermalization.
The kinetic energy spectra of two kinds of photons show the
exponential form which provides us with information on the apparent
temperature in different stage of heavy-ion collision.

The azimuthal asymmetry parameters  have been investigated in
detail. The time evolution of directed flow and elliptic flow of
hard photons exhibits rich structure as the density of
the system oscillates  during the pre-equilibrium and thermalization stage of
reaction system. This structure indicates that the azimuthal
asymmetry evolves with time. A nonzero directed transverse flow
and elliptic flow parameter have been predicted for the direct hard
photons produced by the first channel proton-neutron
bremsstrahlung process in the intermediate energy heavy-ion
collisions. The asymmetry parameters of hard photons are plotted as a function of
rapidity and transverse momentum, which show contrary signs in
comparison with the flow of free protons, i.e. the azimuthal
asymmetry of direct hard photons seems to be anticorrelated to that of 
the corresponding free proton's. Therefore we can expect the
direct hard photon can server as a good probe to nuclear matter
properties in the early stage of HIC.

Different variables dependences are investigated for the hard
photon production and anisotropic flow. We can see that the
absolute value of both directed flow and elliptic flow  of direct
photons and free protons increase with the impact parameter,
indicating that the flow mainly reflects the initial geometric
asymmetry of the collision zone in a given beam energy and revealing a
rich structure of flows with increasing beam energy. For
directed flow, direct photons reach a minimum  of around 40$A$ MeV
where the free protons approach the maximum negative value,
while for elliptic flow, the absolute values of direct photons
show decreasing trends with  increasing beam energy. The EOS
dependence for hard photons at different time indicates  that the
direct photon is not sensitive to the nuclear incompressibility nor to  the symmetry energy. However, for the thermal photon, its
multiplicity increases with nuclear incompressibility. For a
given nuclear incompressibility, the soft symmetry energy favors
thermal photon production.

Finally, we calculated the two-photon correlations, including HBT momentum
correlations and azimuthal correlations. From two-photon HBT
correlations, we can extract  photon source information,
such as intensity interference  and the spatial extent of the emission
source. We also find that two-photon azimuthal correlations
can provide us with a good method for extracting the amplitude of the elliptic flow
parameter $v_{2}$ of hard photons in the experiment.

In light of the present study, we expect that direct photons would
be a very useful probe for exploring nuclear reaction dynamics in
intermediate energy heavy-ion collisions, while the thermal hard
photon can give us some hints on the nuclear EOS, including the
symmetry energy.

\section*{Acknowledgements}

This work is supported partially by  the Knowledge
Innovation Project of Chinese Academy of Sciences under Grant No.
KJCX2-EW-N01, the National Natural Science
Foundation of China under contract No.s 11005140, 11035009,
10979074, 10875160, 10805067 and 10975174.

%\end{CJK*}

\begin{thebibliography}{0}
\bibitem{schutz}Y. Schutz,
    G. Mart\'{\i}nez,
    F.M. Marqu\'es,
    A. Mar\'{\i}n,
    T. Matulewicz,
    R.W. Ostendorf,
    P. Bo\'zek,
    H. Delagrange,
    J. D\'{\i}az,
    M. Franke,
    K.K. Gudima,
    S. Hlav\'a\v{c},
    R. Holzmann,
    P. Lautridou,
    F. Lef\`{e}vre,
    H. L\"ohner,
    W. Mittig
    M. Ploszajczak,
    J.H.G. van Pol,
    J. Qu\'ebert,
    P. Roussel-Chomaz,
    A. Schubert,
    R.H. Siemssen,
    R.S. Simon,
    Z. Sujkowski,
    V.D. Toneev,
    V. Wagner,
    H.W. Wilschut,
    Gy. Wolf,  Nucl. Phys. A {\bf 622}, 404 (1997).

\bibitem{cassing}W. Cassing, V. Metag, U. Mosel, K. Niita, Phys. Rep. {\bf 188}, 363 (1990).

\bibitem{Bona}A. Bonasera, R. Coniglione, and P. Sapienza, Euro. Phys. J. A
{\bf 30}, 47 (2006).

\bibitem{nifenecker}H. Nifenecker and J. A. Pinston,  Annu. Rev. Nucl. Part. Sci.  {\bf 40}, 113 (1990).

\bibitem{wada}R. Wada,  
D. Fabris, K. Hagel, G. Nebbia, Y. Lou, M. Gonin,  J. B. Natowitz,
R. Billerey, B. Cheynis, A. Demeyer, D. Drain, D. Guinet, C. Pastor, L. Vagneron,  K. Zaid,
J. Alarja, A. Giorni, D. Heuer, C. Morand, B. Viano, C. Mazur, C. Ng, S. Leray, R. Lucas, M. Ribrag, and E. Tomasi
Phys. Rev. C {\bf 39}, 497 (1989).

\bibitem{schutz1}Y. Schutz {\it et al.} (TAPS Collaboration),  Nucl. Phys. A {\bf 599}, 97 (1996).

\bibitem{schutz2}Y. Schutz {\it et al.}  (TAPS Collaboration),  Nucl. Phys. A {\bf 630}, 126 (1998).


\bibitem{martinez}G. Martinez, F.M. Marqu\'es, Y. Schutz, Gy. Wolf, J. D\'iaz, M. Franke, S. Hlav\'a\v{c}, R. Holzmann, P. Lautridou,  F. Lef\\`{e}vre,  H. L\"{o}hner, A. Mar\'{\i}n, T. Matulewicz, W. Mittig, R.W. Ostendorf, J.H.G. van Pol, J. Qu\'{e}bertf, P. Roussel-Chomaz, A. Schubert, R.H. Siemssen, R.S. Simon, Z. Sujkowski, V. Wagner, H.W. Wilschut,  Phys. Lett. B {\bf 349}, 23 (1995).

\bibitem{enterria}D. G. d'Enterria, L. Aphecetche, A. Chbihi, H. Delagrange, J. D\'iaz, M. J. van Goethem, M. Hoefman, A. Kugler, H. L\"{o}hner, G. Mart\'{\i}nez, M. J. Mora, R. Ortega, R. W. Ostendorf, S. Schadmand, Y. Schutz, R. H. Siemssen, D. Stracener, P. Tlusty, R. Turrisi, M. Volkerts, V. Wagner, H. W. Wilschut, and N. Yahlali,  Phys. Rev. Lett. {\bf 87}, 022701 (2001).


\bibitem{Phenix} A. Adare {\it et al.} (PHENIX Collaboration), Phys. Rev. Lett. {\bf
104}, 132301 (2010).

\bibitem{Long}J. L. Long, Z. J. He, Y. G. Ma, B. Liu, Phys. Rev. C {\bf 72}, 064907 (2005); Nucl. Phys. A {\bf 766}, 201
(2006).

\bibitem{Liu} F. M. Liu, T. Hirano, K. Werner, and Y. Zhu, Phys. Rev. C {\bf 79}, 014905 (2009).

\bibitem{WA}M. M. Aggarwal {\it et al.} (WA98 Collaboration), Phys. Rev. Lett. {\bf 85}, 3595 (2000).

\bibitem{Bauer} W. Bauer, G. F. Bertsch, W. Cassing, U. Mosel,
 Phys. Rev. C {\bf 34}, 2127 (1986).

\bibitem{Bertsch}G. F. Bertsch, S. Das Gupta,  Phys. Rep. {\bf 160}, 189 (1988).

\bibitem{Jackson}J. D. Jackson, Classical Electrodynamics (Wiley, New York, 1962), p. 733.

\bibitem{Cassing2}W. Cassing, T. Biro, U. Mosel, M. Tohyama, W. Bauer,  Phys. Lett. B {\bf 181}, 217 (1986).

\bibitem{GHLiu}G. H. Liu, Y. G. Ma, X.Z. Cai, D.Q. Fang, W.Q. Shen, W.D. Tian, K. Wang,  Phys. Lett. B {\bf 663}, 312 (2008).

\bibitem{marques}F. M. Marqu\'es, G. Mart\'{\i}nez, Y. Schutz, J. D\'{\i}az, M. Franke, S. Hlav\'{a}\v{c}, R. Holzmann, P. Lautridou, F. Lef\`{e}vre, H. L\"{o}hner, A. Mar\'{\i}n, T. Matulewicz,  W. Mittig, R.W. Ostendorf,  J.H.G. van Pol, J. Qu\'{e}bert, P. Roussel-Chomaz, A. Schubert,  R.H. Siemssen, R.S. Simon, Z. Sujkowski, V. Wagner, H.W. Wilschut, Gy. Wolf,
 Phys. Lett. B {\bf 349}, 30 (1995).



\bibitem{dipole}A. Corsi, O. Wieland, V.L. Kravchuk, A. Bracco, F. Camera, , G. Benzoni, N. Blasi, S. Brambilla, F.C.L. Crespi, A. Giussani, S. Leoni, B. Million, D. Montanari,  A. Moroni, F. Gramegna, A. Lanchais, P. Mastinu, M. Brekiesz, M. Kmiecik, A. Maj, M. Bruno, M. D'Agostino, E. Geraci, j, G. Vannini, S. Barlini, G. Casini, M. Chiari, A. Nannini, A. Ordine, M. Di Toro, C. Rizzo, M. Colonna, V. Barank, Phys. Lett. B {\bf 679}, 197 (2009).

\bibitem{dipole2}V. Baran, C. Rizzo, M. Colonna, M. DiToro, D. Pierroutsakou, Phys. Rev. C {\bf 79}, 021603(R) (2009).


\bibitem{Wu}H. L. Wu, W. D. Tian, Y. G. Ma, X. Z. Cai, J. G. Chen, D. Q. Fang, W. Guo, and H. W. Wang,  Phys. Rev. C {\bf 81}, 047602
(2010).

\bibitem{Bona2_PR} A. Bonasera, F. Gulminelli, J. Molitoris, Phys. Rep. {\bf 243},
1 (1994).


\bibitem{Olli}J. Y. Ollitrault,  Phys. Rev. D {\bf 46}, 229 (1992).

\bibitem{Voloshin} S. Voloshin, Y. Zhang,  Z. Phys. C {\bf 70}, 665 (1996).

\bibitem{Sorge}H. Sorge,  Phys. Lett. B {\bf 402}, 251 (1997);
 Phys. Rev. Lett. {\bf 78} (1997) 2309; {\bf 82}, 2048 (1999).

\bibitem{Danile}P. Danielewicz, R. A. Lacey, P. B. Gossiaux,  C. Pinkenburg, P. Chung, J. M. Alexander, and R. L. McGrath,  Phys. Rev. Lett. {\bf 81}, 2438 (1998).

\bibitem{Ma}Y. G. Ma, W. Q. Shen, J. Feng, Y. Q. Ma,  Phys. Rev. C {\bf 48}, R1492 (1993);  Z. Phys. A {\bf 344}, 469 (1993);  Y. G. Ma, W. Q. Shen, Z. Y. Zhu, Phys. Rev. {C \bf 51}, 1029 (1995); Y.G. Ma, T.Z. Yan, X.Z. Cai, J.G. Chen, D.Q. Fang, W. Guo, G.H. Liu, C.W. Ma, E.J. Ma, W.Q. Shen, Y. Shi, Q.M. Su, W.D. Tian, H.W. Wang, K. Wang, 
 Nucl Phys. A {\bf 787}, 611c (2007).

\bibitem{Ma95}Y. G. Ma and W. Q. Shen, Phys. Rev. C {\bf 51}, 3256
(1995).

\bibitem{Zheng}Y. M. Zheng, C. M. Ko, B. A. Li, and B. Zhang,  Phys. Rev. Lett. {\bf 83}, 2534 (1999).

\bibitem{Gale}D. Persram and C. Gale,  Phys. Rev. C {\bf 65}, 064611 (2002).

\bibitem{INDRA}J. Lukasik {\it et al.} (INDRA-ALDAIN Collaboration),  Phys. Lett. B {\bf 608}, 223 (2004).


\bibitem{Yan}T. Z. Yan, Y. G. Ma, X. Z. Cai, J. G. Chen, D. Q. Fang, W. Guo, C. W. Ma, E. J. Ma, W. Q. Shen, W. D. Tian, K. Wang,  Phys. Lett. B {\bf 638}, 50 (2006).


\bibitem{Chen} J. H. Chen, Y. G. Ma, G. L. Ma, X. Z. Cai, Z. J. He, H. Z. Huang, J. L. Long, W. Q. Shen, C. Zhong, and J. X. Zuo,  Phys. Rev. C {\bf 74},  064902 (2006).

\bibitem{Aggarwal}M. M. Aggarwal {\it et al.} (WA98 Collaboration),  Phys. Rev. Lett. {\bf 93}, 022301 (2004);  Nucl. Phys. A {\bf 762}, 129 (2005) .

\bibitem{Adams}S. S. Adler {\it et al.} (PHENIX Collaobration),  Phys. Rev. Lett. {\bf 96}, 032302 (2006).

\bibitem{Turbide}S. Turbide, C. Gale, R. J. Fries,  Phys. Rev. Lett. {\bf 96}, 032303 (2006).

\bibitem{Barz}H. W. Barz, B. Kampfer,  G. Wolf, W. Bauer, Phys. Rev. C  {\bf 53}, R553 (1996) .

\bibitem{Marques_96}  F. M. Marques, G. Martinez, T. Matulewicz, R. W. Ostendorf, Y. Schutz,  Phys. Rev. C {\bf 54}, 2783 (1996).

\bibitem{Herrmann}V. Herrmann, J. Speth,  K. Nakayama,
Phys. Rev. C {\bf 43}, 394 (1991).



\bibitem{Safkan}Y. Safkan, T. Akdogan, W. A. Franklin, J. L. Matthews, W. M. Schmitt,  V. V. Zelevinsky, 
P. A. M. Gram, T. N. Taddeucci, S. A. Wender, S. F. Pate,  Phys. Rev. C {\bf 75}, 031001 (2007).


\bibitem{Heiselberg}H. Heiselberg and M. Hjorth-Jensen,  Phys. Rep. {\bf 328}, 237 (2000).
\bibitem{LBA}B. A. Li, A. T. Sustich, and B. Zhang,  Phys. Rev. C {\bf  64}, 054604 (2001).
\bibitem{Lane} A. M. Lane, Nucl. Phys.  A {\bf 35}, 676 (1962).

\bibitem{HBT}R. Hanbury-Brown and R.Q. Twiss,  Philos. Mag.  {\bf
45}, 663 (1954).

\bibitem{Goldhaber}G. Goldhaber,  S. Goldhaber, W. Lee,  A. Pais,  Phys. Rev. {\bf
120}, 300 (1960).

\bibitem{Lorstad}B. L\"orstad,  Int. J. Mod. Phys. {\bf A4}, 2861
(1989).

\bibitem{Boal}D.H. Boal, C. K. Gelbke, B. K. Jennings,  Rev. Mod. Phys. {\bf 62}, 553 (1990).

\bibitem{Quebert}J. Qu\'ebert,  Ann. Phys. Fr. {\bf 17}, 99 (1992).

\bibitem{Koonin}S. E. Koonin, Phys. Lett. B {\bf 70}, 43 (1977).
\bibitem{Pratt} S. Pratt, Phys. Rev. Lett. {\bf 53}, 1219 (1984).

\bibitem{Lynch}W. G. Lynch, C. B. Chitwood, M. B. Tsang, D. J. Fields, D. R. Klesch,  C. K. Gelbke,
G. R. Young, T. C. Awes, R. L. Ferguson, F. E. Obenshain, F. Plasil,  R. L. Robinson,
A. D. Panagiotou, Phys. Rev. Lett. {\bf 51}, 1850
(1983).

\bibitem{Marques_RP}F. M. Marques, G. Martinez, T. Matulewicz, R. W. Ostendorf, Y. Schutz, Phys. Rep. {\bf 284}, 91 (1997).

\bibitem{Ma-hbt}Y. B. Wei, Y. G. Ma, W. Q. Shen, G. L. Ma, K. Wang, X. Z. Cai, C. Zhong, W. Guo, J. G. Chen, Phys. Lett. B {\bf
586}, 225  (2004).
\bibitem{Ma-hbt2} Y. G. Ma, Y. B. Wei, W. Q. Shen, X. Z. Cai, J. G. Chen, J. H. Chen, D. Q. Fang, W. Guo, C. W. Ma, G. L. Ma, Q. M. Su, W. D. Tian, K. Wang, T. Z. Yan, C. Zhong, and J. X. Zuo, Phys. Rev. C {\bf 73}, 014604 (2006);  Y. G. Ma, X. Z. Cai, J. G. Chen, D. Q. Fang, W. Guo, G. H. Liu, C. W. Ma, E. J. Ma, W. Q. Shen, Y. Shi, Q. M. Su, W. D. Tian, H. W. Wang, K. Wang, Y. B. Wei, T. Z. Yan, Nucl. Phys. A {\bf 790}, 299c (2007).


\bibitem{Marques2} M. Marqu\'es, R. W. Ostendorf, P. Lautridou, F. Lefevre, T. Matulewicz, W. Mittig, P. Roussel-Chomaz, Y. Schutz, J. Qu\'{e}bert, J. D\'{\i}az, A. Mar\'{\i}n,  G. Mart\'{\i}nez, R. Holzmann, S. Hl\'{a}va\v{c}, A. Schubert, R. S. Simon,  V. Wagner, H. L\"{o}hner, J. H. G. van Pol, R. H. Siemssen, H. W. Wilschut, M. Franke, Z. Sujkowski, Phys. Rev. Lett. {\bf 73}, 34 (1994); F. M. Marqu\'es, G. Martinez, T. Matulewicz, R. W. Ostendorf, Y. Schutz, Phys. Rev. C {\bf 54},  2783 (1996).

\bibitem{Wang}S. Wang,  Y. Z. Jiang, Y. M. Liu, D. Keane, D. Beavis, S. Y. Chu, S. Y. Fung, and M. Vient, C. Hartnack, H. St\"{o}cker,  Phys. Rev. C {\bf 44}, 1091 (1991).

\bibitem{Lacey}R. A. Lacey, A. Elmaani, J. Lauret, T. Li, W. Bauer, D. Craig, M. Cronqvist, E. Gualtieri, S. Hannuschke, T. Reposeur, A. Vander Molen, G. D. Westfall, W. K. Wilson, J. S. Winfield, J. Yee, S. Yennello, A. Nadasen, R. S. Tickle, and E. Norbeck,  Phys. Rev. Lett. {\bf 70}, 1224
(1993).



\end{thebibliography}
\end{document}